\documentclass[aps,pra,twocolumn,superscriptaddress,nofootinbib]{revtex4-1}
\usepackage{amsmath}
\usepackage{amssymb}
\usepackage{graphicx}
\usepackage{mathrsfs}
\usepackage{ntheorem}
\usepackage{mathtools}
\usepackage{enumerate}
\usepackage{enumitem}
\usepackage{times,txfonts}
\usepackage{subfigure}
\usepackage{upgreek}
\usepackage{tikz}
\newcommand{\ket}[1]{|#1\rangle}
\newcommand{\bra}[1]{\langle #1|}
\newcommand{\Tr}{\mathrm{Tr}}

\newcommand{\abs}[1]{\lvert #1\rvert}

\def\CC{{\rm\kern.24em \vrule width.04em height1.46ex depth-.07ex \kern-.30em C}}
\def\RR{{\rm\kern.24em \vrule width.04em height1.46ex depth-.07ex
\kern-.30em R}}
\def\P{{\rm I\kern-.25em P}}

\begin{document}
\title{Catalyst-Assisted Probabilistic Coherence Distillation for Mixed States}
\author{C. L. Liu}
\affiliation{Institute of Physics, Beijing National Laboratory for Condensed Matter Physics, Chinese Academy of Sciences, Beijing 100190, China}
\author{D. L. Zhou}
\email{zhoudl72@iphy.ac.cn}
\affiliation{Institute
  of Physics, Beijing National Laboratory for Condensed Matter
  Physics, Chinese Academy of Sciences, Beijing 100190, China}
\affiliation{School of Physical Sciences, University of Chinese
  Academy of Sciences, Beijing 100049, China}
\affiliation{CAS Central of Excellence in Topological Quantum
  Computation, Beijing 100190, China}
\affiliation{Songshan Lake Materials Laboratory, Dongguan, Guangdong
  523808, China}
\date{\today}
\begin{abstract}
The remarkable phenomenon of catalyst tells us that adding a catalyst could help state transformation. In this paper, we consider the problem of catalyst-assisted probabilistic coherence distillation for mixed states under strictly incoherent operations.  To this end, we first present the necessary and sufficient conditions for distilling a target pure coherent state from an initial mixed state via stochastic strictly incoherent operations and  the maximal probability of obtaining the target pure state from the initial state. With the help of these results, we present the necessary and sufficient conditions for the existence of a catalyst that increases the maximal transformation probability.
\end{abstract}
\maketitle

\section{Introduction}
Quantum coherence is a fundamental quantum resource of quantum physics, describing the capability of a quantum state to exhibit quantum interference phenomena. It is an essential component in quantum information processing \cite{Nielsen}, and plays a central role in emergent fields, such as quantum computing \cite{Shor,Grover}, quantum cryptography \cite{Bennett},  quantum metrology \cite{Giovannetti,Giovannetti1}, and quantum biology \cite{Lambert}. Hence, the resource theory of coherence has attracted a growing interest due to the development of quantum information science in recent years  \cite{Aberg1,Baumgratz,Levi,Streltsov,Fan}. For a quantum resource theory, there are two fundamental ingredients:  free states and free operations \cite{Horodecki,Chitambar,Brandao}.  With regards to the resource theory of coherence, the free states are quantum states which are diagonal in a prefixed reference basis. There is no general consensus on the set of free operations in the resource theory of coherence. Based on various physical and mathematical considerations, several free operations of coherence were presented \cite{Streltsov}. Here, we focus our attention on the strictly incoherent operations which was given in Ref. \cite{Winter}. It was shown that the strictly incoherent operations neither create nor use coherence and have a physical interpretation in terms of interferometry in Ref. \cite{Yadin}. Thus, the strictly incoherent operations are a physically well-motivated set of free operations for coherence and a strong candidate for free operations.

When we perform a quantum information processing task, it is usually the pure coherent states playing the central role \cite{Nielsen}. Unfortunately, as a quantum system is unavoidably affected by noise, a pure state easily becomes a mixed state. Thus, a central problem in the resource theory of coherence is the coherence distillation, i.e., the process that extracts target pure coherent states from initial mixed states via incoherent operations.  Recently, this problem has generated a great deal of interest \cite{Winter,Liu,Yuan,Fang,Lami,Lami1,Regula,Regula1,Chitambar1,Zhao,Liu2,Bu1}. The coherence distillation of mixed states via various incoherent operations was studied in the asymptotic limit in Refs. \cite{Winter,Lami1,Zhao}. Another scheme, the scheme of one-shot coherence distillation, i.e., the process that extracts pure coherent states from mixed states via various incoherent operations in the nonasymptotic limit, was proposed in Refs. \cite{Liu,Yuan,Fang,Lami,Regula,Regula1,Chitambar1,Zhao, Liu2,Bu1}. When we consider the strictly incoherent operations, it shows that, for a class of mixed states, it can never be transformed into a target pure state via strictly incoherent operations with certainty or with high probability \cite{Liu, Liu2, Lami1, Zhao}.  Thus, an important problem is that how to increase the transformation probability in this situation.

Inspired by the catalyst-assisted transformations in the resource theory of quantum entanglement \cite{Jonathan,Turgut,Klimesh,Feng}, catalyst-assisted coherence transformations for pure states \cite{Du1,Bu} was proposed as an efficient method to increase the transformation probability of coherence distillation. Specifically, if the initial state and the target state are all pure states, they show that an appropriately chosen catalyst can increase the maximal transformation probability and they present the necessary and sufficient conditions for the existence of a catalyst that can increase the maximal transformation probability. However, in practical applications, because a quantum system is unavoidably affected by its environment, we would expect to deal with the coherence distillation of mixed states rather than with pure states. Therefore, an immediate question arises: With the help of coherence catalysts, can we increase maximal transformation probability of coherence distillation for the initial states being mixed states? In other words, for the initial states being mixed states, what are the necessary and sufficient conditions for the existence of a catalyst that can increase the maximal transformation probability of coherence distillation?

In this work, we address the above question by considering the problem of catalyst-assisted probabilistic coherence distillation for mixed states under strictly incoherent operations.  To solve this problem, we divide our discussions into three steps. In the first step, we present the necessary and sufficient conditions for distilling a target pure coherent state from an initial mixed state via stochastic strictly incoherent operations. In the second step, we present the maximal probability of obtaining the target pure state from the initial state. In the last step, we present the necessary and sufficient conditions for the existence of a catalyst that can increase the maximal transformation probability.

This paper is organized as follows. In Sec.~II, we recall some notions of the quantum resource theory of coherence. In Sec.~III,  we present the necessary and sufficient conditions for the existence of a catalyst that can increase the maximal transformation probability. Section IV is our conclusions.

\section{Resource theory of coherence}
Let $\mathcal {H}$ be the Hilbert space of a $d$-dimensional quantum system. A particular basis of $\mathcal {H}$ is denoted as $\{\ket{i}, ~i=1,2,\cdot\cdot\cdot,d\}$, which is chosen according to the physical problem under discussion. Coherence of a state is then measured based on the basis chosen. Specifically, a state is said to be incoherent if it is diagonal in the basis. Any state which cannot be written as a diagonal matrix is defined as a coherent state. For a pure state $\ket{\varphi}$, we will denote $\ket{\varphi}\bra{\varphi}$ as $\varphi$, i.e., $\varphi:=\ket{\varphi}\bra{\varphi}$.

Next, let us recall strictly incoherent operations \cite{Winter, Yadin}. A strictly incoherent operation is a completely positive trace-preserving map, expressed as
\begin{eqnarray}
\Lambda(\rho)=\sum_n K_n\rho K_n^\dagger,
\end{eqnarray}
where the Kraus operators $K_n$ satisfy not only  $\sum_n K_n^\dagger K_n= I$ but also $K_n\mathcal{I}K_n^\dagger\subset \mathcal{I}$ and $K_n^\dagger\mathcal{I}K_n\subset\mathcal{I}$ for $K_n$, i.e., each $K_n$ as well $K_n^\dagger$ maps an incoherent state to an incoherent state. Here, $\mathcal{I}$ represents the set of incoherent states.
There is at most one nonzero element in each column (row) of $K_n$, and such a $K_n$ is called a strictly incoherent Kraus operator. With this definition, it is elementary to show that a projector is an incoherent operator if and only if it has the form $\mathbb{P}_\alpha=\sum_{i\in\alpha}\ket{i}\bra{i}$ with $\alpha\subset\{1,...,d\}$. In what follows, we will denote $\mathbb{P}_\alpha$ as strictly incoherent projective operators. The dephasing map, which we will denote it as $\Delta(\cdot)$, is defined as $\Delta(\rho)=\sum_{i=1}^{d}\ket{i}\bra{i}\rho\ket{i}\bra{i}$.

With the aid of strictly incoherent operations, we then recall the notion of stochastic strictly incoherent operations \cite{Liu}. A stochastic strictly incoherent operation is constructed by a subset of strictly incoherent Kraus operators. Without loss of generality, we denote the subset as $\{K_{1},K_{2},\dots, K_{L}\}$. Otherwise, we may renumber the subscripts of these Kraus operators. Then, a stochastic strictly incoherent operation, denoted $\Lambda_s(\rho)$, is defined by
\begin{equation}
\Lambda_s(\rho)=\frac{\sum_{n=1}^L K_{n}\rho K_{n}^{\dagger}}{\Tr(\sum_{n=1}^LK_{n}\rho K_{n}^{\dagger})},
\label{lams}
\end{equation}
where $\{K_{1},K_{2},\dots, K_{L}\}$ satisfies $\sum_{n=1}^L K_{n}^{\dagger}K_{n}\leq I$. Clearly, the state $\Lambda_s(\rho)$ is obtained with probability $P=\Tr(\sum_{n=1}^LK_{n}\rho K_{n}^{\dagger})$ under a stochastic strictly incoherent operation $\Lambda_s$, while the state $\Lambda(\rho)$ is fully deterministic under a strictly incoherent operation $\Lambda$.

A functional $C$ can be taken as a measure of coherence if it satisfies the four postulates \cite{Baumgratz,Yadin,Streltsov,Fan}: (C1) the coherence being zero (positive) for incoherent states (all other states); (C2) the monotonicity of coherence under strictly incoherent operations; (C3) the monotonicity of coherence under selective measurements on average; and (C4) the nonincreasing of coherence under mixing of quantum states. In accordance with the general criterion, several coherence measures have been put forward. Out of them, we recall the following coherence measures, which are considered in this work. The coherence rank $C_r$ \cite{Winter,Killoran} of a pure state (not necessarily normalized), $\ket{\varphi}=\sum_{i=1}^R\varphi_i\ket{i}$ with $\varphi_i\neq 0$, is defined as the number of terms with $\varphi_i\neq0$, i.e., $C_r(\varphi)=R$. For an arbitrary pure state $\ket{\varphi}=\sum_{i=1}^{d}\varphi_i\ket{i}$, we define $
C_l(\ket{\varphi})=\sum_{i=l}^{d}\abs{\varphi_i}^2$,
with $l=1,2,\cdots,d$ and $\abs{\varphi_1}\geq\cdots\geq\abs{\varphi_{d}}$.
These measures were put forward in Ref. \cite{Du1}.

\section{Catalyst-Assisted Probabilistic Coherence Distillation for Mixed States}

We start by specifying the notion of catalyst-assisted probabilistic transformation under stochastic strictly incoherent operations. For a given initial state $\rho$ and a target state $\varphi$, we denote the maximal probability of obtaining the state $\varphi$ from $\rho$ by means of  stochastic strictly incoherent operations as $P_{\max}(\rho\to\varphi)$, i.e.,
\begin{eqnarray}
P_{\max}(\rho\to\varphi)=\max_{\Lambda_s}\Tr\left(\Lambda_s(\rho)\right),
\end{eqnarray}
with $\Lambda_s(\rho)\propto\varphi$. With the notion of  $P_{\max}(\rho\to\varphi)$, we say that the transformation from $\rho$ to $\varphi$ can be enhanced by using a catalyst if there exists some catalyst $c$ such that
\begin{eqnarray}
P_{\text{max}}(\rho\otimes c \to \varphi\otimes c)>P_{\text{max}}(\rho \to \varphi).
\end{eqnarray}
Otherwise, we say that the maximal transformation probability $P_{\max}(\rho\to\varphi)$ cannot be increased via catalysts.

To give the necessary and sufficient conditions for the existence of a catalyst that can increase the maximal transformation probability, we present our result as the following three steps.

First, we present the necessary and sufficient conditions for distilling a target pure coherent state from an initial mixed state via stochastic strictly incoherent operations. 

We state the condition as the following Theorem.

\emph{Theorem 1.} One can distill a target pure coherent state $\varphi$ with its coherence rank $C_r(\varphi)=m\leq d$  from an initial coherent state $\rho$ via a stochastic strictly incoherent operation $\Lambda_s$ if and only if there exists an incoherent projective operator $\mathbb{P}$ such that
\begin{eqnarray}
\frac{\mathbb{P}\rho\mathbb{P}}{\Tr(\mathbb{P}\rho\mathbb{P})}=\psi, \label{theorem1}
\end{eqnarray}
with the coherence rank of $\psi$ being $n(\geq m)$.

We now prove the theorem.

For the \emph{if} part, we have to show that if the state $\rho$ fulfills the condition in the theorem, i.e., Eq. (\ref{theorem1}), then we can always construct an explicit stochastic strictly incoherent operation such that $\Lambda_s(\rho)=\varphi$.

To this end,  by using the condition in Eq. (\ref{theorem1}), we can obtain an incoherent projective operator $\mathbb{P}$ such that $\frac{\mathbb{P}\rho\mathbb{P}}{\Tr(\mathbb{P}\rho\mathbb{P})}=\psi$ with the coherence rank of $\psi$
being equal to or greater than that of $\varphi$. Without loss of generality, we assume that there is $\ket{\varphi}=\sum_{i=1}^m\varphi_i\ket{i}$ with $\abs{\varphi_1}\geq\abs{\varphi_2}\geq\cdots\geq\abs{\varphi_m}>0$. Furthermore, we
assume that $\ket{\psi}=U\sum_{i=1}^n\psi_i\ket{i}$, where $U$ is a permutation matrix such that there is $\abs{\psi_1}\geq\abs{\psi_2}\geq\cdots\abs{\psi_n}>0$.
Then, the stochastic strictly incoherent operation $\Lambda_s$ can be chosen as
\begin{eqnarray}
\Lambda_s(\rho)=\frac{K\rho K^\dag}{\Tr(K\rho K^\dag)},
\end{eqnarray}
where
\begin{eqnarray}
K=k~\text{diag}\left(a_1,a_2,\cdots,a_n\right)U^\dag\mathbb{P},\nonumber
\end{eqnarray}
with $a_i=\frac{\varphi_i}{\psi_i}$ for all $i=1,\cdots,n$ and $k$ being a complex number for guaranteeing $\abs{k}\leq\frac{1}{\max_i\{\abs{a_i}\}}$. It is straightforward to show that $\Lambda_s(\rho)=\Tr[\Lambda_s(\rho)]\varphi$. This completes the \emph{if} part of the theorem.

Next, we show the \emph{only if} part of the theorem, i.e.; if we can distill $\varphi$ from $\rho$ via a stochastic strictly incoherent operation, then
there exists an incoherent projective operator $\mathbb{P}$ such that $\frac{\mathbb{P}\rho\mathbb{P}}{\Tr(\mathbb{P}\rho\mathbb{P})}=\psi$
with the coherence rank of $\psi$ being equal to or larger than that of $\varphi$.

First, we show that when we want to judge whether there exists a stochastic strictly incoherent operation such that $\Lambda_s(\rho)=\varphi$, we only need to consider the stochastic strictly incoherent operation with the form of
\begin{eqnarray}
\Lambda_s^1(\rho)=\frac{K\rho K^\dag}{\Tr(K\rho K^\dag)}. \label{Step1}
\end{eqnarray}

To this end, on the one hand, we assume that we can distill a  given pure coherent state $\varphi$ from $\rho$ via a stochastic strictly incoherent operation  $\Lambda_s$, i.e.,  
\begin{equation}\nonumber\Lambda_s(\rho)=\frac{\sum_{n=1}^L K_{n}\rho K_{n}^{\dagger}}{\Tr(\sum_{n=1}^LK_{n}\rho K_{n}^{\dagger})}=\varphi.\end{equation} Then,
since pure states are extreme points of the set of states, there must be
$\frac{K_{n}\rho K_{n}^{\dagger}}{\Tr(K_{n}\rho K_{n}^{\dagger})}=\varphi$
for all $n=1,...,L$.
On the other hand, we note that  \begin{equation}\nonumber\Lambda_s^1(\rho)=\frac{K_n(\rho)K_n^\dag}{\Tr(K_n(\rho)K_n^\dag)}\end{equation} is also a stochastic strictly incoherent operation. Thus, we obtain that if we can distill $\varphi$ from $\rho$ via $\Lambda_s^1(\rho)$, then there exists a stochastic strictly incoherent operation such that $\Lambda_s(\rho)=\varphi$.

Second, we show that if we can distill a pure state $\varphi$ from a state $\rho$ via a stochastic strictly incoherent operation, there must be
\begin{eqnarray}
\frac{\mathbb{P}\rho\mathbb{P}}{\Tr(\mathbb{P}\rho\mathbb{P})}=\psi,\label{Step2}
\end{eqnarray}
with the coherence rank of $\psi$ being equal to or larger than that of $\varphi$.

To this end, let us recall the structure of the strictly incoherent Kraus operator \cite{Liu2}. From the definition of the strictly incoherent Kraus operator, we obtain that any strictly incoherent Kraus operator $K$ can always be decomposed into $K=P_\pi K_\Delta\mathbb{P}$,
where the operator $P_\pi$ is a permutation matrix, $K_\Delta=\text{diag}(a_1,...,a_n,0,0,...)$ is a diagonal matrix with $a_i\neq0$, and $\mathbb{P}$ is a projective operator corresponding to $K_\Delta$, i.e.,
$\mathbb{P}=\text{diag}(1,...,1,0,0,...)$. By using $K=P_\pi K_\Delta\mathbb{P}$ and Eq. (\ref{Step1}), we then obtain that $
\frac{P_\pi K_\Delta\mathbb{P}\rho\mathbb{P}K_\Delta^\dag P_\pi^\dag}{\Tr(P_\pi K_\Delta\mathbb{P}\rho\mathbb{P}K_\Delta^\dag P_\pi^\dag)}=\varphi. $
Without loss of generality, we assume that there is $
\mathbb{P}\rho\mathbb{P}=p\sum_\alpha p_\alpha\varphi_\alpha$, with $p=\Tr(\mathbb{P}\rho\mathbb{P})$. Thus, we immediately obtain that
\begin{eqnarray}
\frac{\sum_\alpha p_\alpha P_\pi K_\Delta\varphi_\alpha K_\Delta^\dag P_\pi^\dag}{\Tr(\sum_\alpha p_\alpha P_\pi K_\Delta\varphi_\alpha K_\Delta^\dag P_\pi^\dag)}=\varphi.\nonumber
\end{eqnarray}
Again, by using the fact that pure states are extreme points of the set of states, we then get that
\begin{eqnarray}
\frac{P_\pi K_\Delta\varphi_\alpha K_\Delta^\dag P_\pi^t}{\Tr(P_\pi K_\Delta\varphi_\alpha K_\Delta^\dag P_\pi^t)}=\varphi~\text{or}~\textbf{0}, \label{ensemble}
\end{eqnarray}
for all $\varphi_\alpha$, where $\textbf{0}$ is a square matrix with all its elements being $0$. Here, we should note that $P_\pi K_\Delta\varphi_\alpha K_\Delta^\dag P_\pi^t$ cannot be $\textbf{0}$ at the same time. This means that there is
$\frac{\mathbb{P}\rho\mathbb{P}}{\Tr(\mathbb{P}\rho\mathbb{P})}=\psi$. Next, by using the result presented in Refs. \cite{Du,Du1,Zhu,Winter}, we can obtain that a pure state $\psi$ can be transformed into a given pure coherent state $\varphi$ via a stochastic strictly incoherent operation if and only if the coherence rank of  $\psi$ is equal to or greater than that of $\varphi$. By using this result and the fact that $\Lambda_s^1(\rho)$ is a strictly incoherent operation, we immediately obtain that the coherence rank of $\psi$ is equal to or larger than that of $\varphi$.

This completes the \emph{only if} part of the theorem.  ~~~~~~~~~~~~~~~~~~~ $\blacksquare$

\emph{Second, we present the maximal probability of obtaining the target pure state $\varphi$ from the initial state $\rho$ by using stochastic strictly incoherent operations.}

Before going further, we give the definition of pure coherent-state subspace. If there is an incoherent projector $\mathbb{P}$ such that $\frac{\mathbb{P_\mu}\rho\mathbb{P_\mu}}{\Tr(\mathbb{P_\mu}\rho\mathbb{P_\mu})}=\psi_\mu$ with the coherence rank of $\psi_\mu$ being $n$, then we say that $\rho$ has an $n$-dimensional pure coherent-state subspace.  Specifically,  the maximally pure coherent-state subspaces of $\rho$ means that the coherence rank of $\psi_\mu$, i.e., the rank of the corresponding $\mathbb{P}$, cannot be enlarged.

From \emph{Theorem 1}, to obtain $P_{\max}(\rho\to\varphi)$, we should solve the problem of how to identify the pure coherent-state subspaces of a mixed state. We can solve this problem with the aid of
\begin{eqnarray}
  \mathcal{A}=(\Delta\rho)^{-\frac12}\abs{\rho}(\Delta\rho)^{-\frac12}\label{A},
\end{eqnarray}
which was given in Ref. \cite{Liu}.
Here, for $\rho=\sum_{ij}\rho_{ij}\ket{i}\bra{j}$, $|\rho|=\sum_{ij}|\rho_{ij}|\ket{i}\bra{j}$ and
$(\Delta\rho)^{-\frac12}$ is a diagonal matrix with elements
$$(\Delta\rho)^{-\frac12}_{ii}= \left\{
  \begin{array}{ll}
     \rho_{ii}^{-\frac12}, &\text{if} ~ \rho_{ii}\neq0;\\
    0,&\text{if}~ \rho_{ii}= 0.
  \end{array}\right.$$
It is straightforward to show that all the elements of $\mathcal{A}$ are $1$ if and only if $\rho$ is a pure state \cite{Liu}.  From this, we obtain that, if there is an $n$-dimensional principal submatrices $\mathcal{A}_\mu$ of $\mathcal{A}$ with all its elements being $1$, then the corresponding subspace of $\rho$ is an $n$-dimensional pure coherent-state subspace. By using this result, one can easily identify the pure coherent-state subspaces of $\rho$. We denote the principal submatrices $\mathcal{A}_\mu$ corresponding to the maximally pure coherent-state subspaces as the maximally dimensional principal submatrices of $\mathcal{A}$. Let the corresponding Hilbert subspaces of principal submatrices $\mathcal{A}_\mu$ be $\mathcal{H}_\mu$, which is spanned by
$\{\ket{i_\mu^1},\ket{i_\mu^2},\cdots,\ket{i_\mu^n}\}\subset\{\ket{1}, \ket{2},\cdots,\ket{d}\}$. Then, the corresponding incoherent projective
operators are
\begin{eqnarray}
  \mathbb{P}_\mu=\ket{i_\mu^1}\bra{i_\mu^1}+\ket{i_\mu^2}\bra{i_\mu^2}+\cdots+\ket{i_\mu^{n}}\bra{i_\mu^{n}}.\nonumber
\end{eqnarray}
Performing $\{\mathbb{P}_\mu\}$ on the state $\rho$, we  obtain
$\{\psi_\mu\}$, i.e.,
\begin{eqnarray}
  \frac{\mathbb{P}_\mu\rho\mathbb{P}_\mu}{\Tr(\mathbb{P}_\mu\rho\mathbb{P}_ \mu)}
  =\psi_\mu.\nonumber
\end{eqnarray}

Equipped with the above notions and Theorem 1, we can present the following Theorem 2.

\emph{Theorem 2.} The maximal probability of obtaining a target coherent state $\varphi$ from an initial state $\rho$ via stochastic strictly incoherent operations is
\begin{eqnarray}
P_{\max}(\rho\to\varphi)=\sum_\mu p_\mu\min_l\frac{C_l(\psi_\mu)}{C_l(\varphi)},
\end{eqnarray}
where $\psi_\mu$ correspond to the maximally pure coherent-state subspaces of $\rho$.

We now prove the theorem.

From \emph{Theorem 1}, we obtain that if we want to get a target coherent state $\varphi$ from an initial state $\rho$ via stochastic strictly incoherent operations, we only need to consider the pure coherent-state subspaces of $\rho$.

First, we show that, to obtain $P_{\max}(\rho\to\varphi)$, we only need to study the maximal pure coherent-state subspace of $\rho$.

To this end, according to the result in Refs. \cite{Vidal,Du,Zhu,Chitambar2}, we can get that
\begin{eqnarray}
P_{\max}(\psi_\mu\to\varphi)=\min_l\frac{C_l(\psi_\mu)}{C_l(\varphi)}. \label{probability}
\end{eqnarray}
 Then, for all $\psi$, $\psi^\prime\in\mathcal{H}$ and $p\in[0,1]$, we may assume that $1\leq l\leq 2n$ and
\begin{eqnarray}
P_{\max}\left(p\psi\oplus(1-p)\psi^\prime\to\varphi\right)=\frac{C_l(p\psi\oplus(1-p)\psi^\prime)}{C_l(\varphi)}\nonumber
\end{eqnarray}
Then, there exist $l_\psi\leq n$ and $l_{\psi^\prime}\leq n$ such that
$C_l(p\psi\oplus(1-p)\psi^\prime)=pC_{l_\psi}(\psi)+(1-p)C_{l_{\psi^\prime}}(\psi^\prime)$.
Thus, there is
\begin{eqnarray}
&&P_{\max}\left(p\psi\oplus(1-p)\psi^\prime \to\varphi\right)=\frac{C_l\left(p \psi\oplus(1-p) \psi^\prime\right)}{C_l(\varphi)}\nonumber\\
&&=\frac{pC_{l_\psi}(\psi)+(1-p)C_{l_{\psi^\prime}}(\psi^\prime)}{C_l(\varphi)}\nonumber\\
&&=\frac{pC_{l}(\psi\oplus 0)+(1-p)C_l(\psi^\prime\oplus0)}{C_l(\varphi)}\nonumber\\
&&\geq pP_{\max}(\psi\to\varphi)+(1-p)P_{\max}(\psi^\prime\to\varphi),
\nonumber\end{eqnarray}
where $\psi\oplus 0$ and $\psi^\prime\oplus0$ means that we append extra zeros to obtain $l$-dimensional matrices, respectively.

Second, we show that the maximal transformation probability is
\begin{eqnarray}
P_{\max}(\rho\to\varphi)=\sum_\mu p_\mu\min_l\frac{C_l(\psi_\mu)}{C_l(\varphi)}.\nonumber
\end{eqnarray}

From \emph{Theorem 1} and the above discussions, we obtain that
\begin{eqnarray}
P_{\max}(\rho\to\varphi)=\sum_\mu p_\mu P_{\max}(\psi_\mu\to\varphi),\nonumber
\end{eqnarray}
where $p_\mu=\Tr(\mathbb{P}_{\mu}\rho\mathbb{P}_ {\mu})$. Then, from Eq. (\ref{probability}),
we immediately obtain
\begin{eqnarray}
P_{\max}(\rho\to\varphi)=\sum_\mu p_\mu\min_l\frac{C_l(\psi_\mu)}{C_l(\varphi)}.
\end{eqnarray}
This completes the proof of the theorem.

By the way, we would like to point that the results of \emph{Theorem 1} and \emph{Theorem 2} are a generalization of the result in Ref. \cite{Liu2}, which provided the conditions for the transformation from $\rho$ into $\varphi$ with certainty. Furthermore, we note that the phenomenon of bound coherence under strictly incoherent operations was uncovered in Refs. \cite{Zhao,Lami,Lami1}, i.e., there are coherent states from which no coherence can be distilled via strictly incoherent operations in the asymptotic regime. The necessary and sufficient conditions for a state being bound state was presented in Refs. \cite{Lami,Lami1}. Their result shows that a state is a bound state if and only if it cannot contain any rank-one submatrix. From the result presented in \emph{Theorem 1}, we obtain a similar result when only one-copy of the state is supplied:

\emph{Corollary 1.} We can distill an arbitrary pure coherent state  from a mixed state $\rho$  via stochastic strictly incoherent operations if and only if $\rho$ contains at least one rank-$2$  pure coherent state subspace.~~~~~~~~~~~~~~~~~ ~~~~~~~~~~~~~~~~~~~~~~~~~~~~~~~~~~~~~~~~~~~~~~~~~~~~~~~~~~~ $\blacksquare$

\emph{Finally, we present the necessary and sufficient conditions for the existence of a catalyst that can increase the maximal transformation probability.}

To present the necessary and sufficient conditions for $P_{\text{max}}(\rho\otimes c \to \varphi\otimes c)>P_{\text{max}}(\rho \to \varphi)$, we need the following lemma, which was presented in Refs. \cite{Bu, Feng}.

\emph{Lemma 1.}---Let $\textbf{p}$ and $\textbf{q}$ be two  $d$-dimensional probability  distributions arranged in nonincreasing order. Then,   there exists a probability distribution $\textbf{c}$ such that $P_{\text{max}}(\textbf{p}\otimes\textbf{c}\to\textbf{p}\otimes\textbf{c})>P_{\max}(\textbf{p}\to\textbf{q})$ if and only if there are $P_{\max}(\textbf{p}\to\textbf{q})<\min\{\frac{p_d}{q_d},1\}$.

With the  \emph{Lemma 1}and \emph{Theorems 1}, and \emph{2},  we immediately obtain the following theorem which provides the necessary and sufficient conditions for the enhancement of  $P_{\max}(\rho\to\varphi)$:

\emph{Theorem 3.}---Suppose the states corresponding to maximally dimensional pure subspaces of $\rho$ are $\psi_\mu$ and, without loss of generality, $\ket{\psi_\mu}=\sum_{i=1}^{n_1}\psi_i^\mu\ket{i}$ with $\abs{\psi_1^\mu}\geq\abs{\psi_2^\mu}\geq\cdots\geq\abs{\psi_{n_1}^\mu}>0$ for all $\mu$ and $\ket{\varphi}=\sum_{i=1}^{n_2}\varphi_i\ket{i}$ with $\abs{\varphi_1}\geq\abs{\varphi_2}\geq\cdots\geq\abs{\varphi_{n_2}}>0$, respectively.  Then $P_{\text{max}}(\rho\otimes c \to \varphi\otimes c)>P_{\text{max}}(\rho \to \varphi)$ if and only if there exist at least one $\psi_\mu$ such that
\begin{eqnarray}
P_{\max}(\psi_\mu\to\varphi)<\min\{\frac{\psi_n^\mu}{\varphi_n},1\},
\end{eqnarray}
where $n=\text{max}\{n_1,n_2\}$. In particular, we consider the catalyst-assisted deterministic coherence transformation. For a given initial state $\rho$ and a target state $\varphi$, we say that the transformation from $\rho$ into $\varphi$ can be catalyzed if there exists some catalyst $c$ such that
\begin{eqnarray}
P_{\text{max}}(\rho\otimes c \to \varphi\otimes c)=1,
\end{eqnarray}
with $P_{\max}(\rho\to\varphi)<1$. To present the necessary and sufficient conditions for the catalytic coherence transformations between the initial state $\rho$ and the target state $\varphi$, we need the following lemma, which was presented in Refs. \cite{Bu, Klimesh,Turgut}.

\emph{Lemma 2.}---Let $\textbf{p}$ and $\textbf{q}$ be two  $d$-dimensional probability  distributions arranged in nonincreasing order with $\textbf{p}$ having nonzero elements. Then,  there exists a probability distribution $\textbf{c}$ such that $\textbf{p}\otimes\textbf{c}\prec\textbf{p}\otimes\textbf{c}$, where $\textbf{p}\prec\textbf{q}$ means that $\sum_{i=1}^lp_i^\downarrow\leq\sum_{i=1}^lq_i^\downarrow$, for all $1\leq l\leq n$ if and only if there are $A_\alpha(\textbf{p})>A_\alpha(\textbf{q}) ~\text{for}~\alpha\in(-\infty,1)$, $A_\alpha(\textbf{p})<A_\alpha(\textbf{q}) ~\text{for}~\alpha\in(1,\infty)$, $S(\textbf{p})>S(\textbf{q})$,
where $A_\alpha(\textbf{p}):=(\frac1d\sum_{i=1}^dp_i^\alpha)^{1/\alpha}$, $S(\textbf{p})=-\sum_{i=1}^dp_i\ln p_i$, and $A_0(\textbf{p}):=(\prod p_i)^{1/d}$.

With the  \emph{Lemma 2}, \emph{Theorems 1}, and \emph{2}, we immediately obtain the following theorem which provides the necessary and sufficient conditions for the deterministically catalytic coherence transformations between the initial state $\rho$ and the target state $\varphi$.

\emph{Theorem 4.}---For an initial state $\rho$ with  the state corresponding to maximally dimensional pure subspaces of $\rho$ being $\psi_\mu$  and a target state $\varphi$, if $P_{\max}(\rho\to\varphi)<1$, then there exists a catalyst $c$ such that $P_{\max}(\rho\otimes c\to\varphi\otimes c)=1$
if and only if there are
\begin{eqnarray}
A_\alpha(\overrightarrow{\Delta\psi_\mu})&&>A_\alpha(\overrightarrow{\Delta\varphi}) ~\text{for}~\alpha\in(-\infty,1);\nonumber\\
A_\alpha(\overrightarrow{\Delta\psi_\mu})&&<A_\alpha(\overrightarrow{\Delta\varphi}) ~\text{for}~\alpha\in(1,\infty);\nonumber\\
	S(\overrightarrow{\Delta\psi_\mu})&&>S(\overrightarrow{\Delta\varphi}),
\end{eqnarray}
 for all $\mu$. Here,  for the pure states $\ket{\psi_\mu}=\sum_{i=1}^n\psi_i^\mu\ket{i}$, $\overrightarrow{\Delta\psi_\mu}$ are defined as $(\abs{\psi^\mu_1}^2,\abs{\psi^\mu_2}^2,\cdots,\abs{\psi^\mu_n}^2)$. ~~~~~~~~~~~~~~~~~~~~~~~~~~~~~~~~~~~~~~~~~~~~~~~~~~~~ $\blacksquare$

\section{Conclusions and Remarks}

To summarize, we have considered the problem of catalyst-assisted probabilistic coherence distillation for mixed states under strictly incoherent operations.  Our main findings are presented as three theorems.  Theorem 1 presents the necessary and sufficient conditions for distilling a target pure coherent state from an initial mixed state via stochastic strictly incoherent operations.  Theorem 2 presents the maximal probability of obtaining the target pure state from the initial state. With the help of these results, we have presented the necessary and sufficient conditions for the existence of a catalyst that can increase the maximal transformation probability in Theorem 5. In passing, there are many open problems to which Theorem 5 may be of relevance. It would be of great interest to determine the maximally achievable transformation probability by the presence of any catalyst state, which is left as an open issue for further investigation.

\section*{Acknowlegements}

We acknowledge fruitful discussions with Prof. Dian-Min Tong and Xiao-Dong Yu. The authors acknowledge financial supported from NSF of China (Grant No.11775300), the National Key Research and Development Program of China (2016YFA0300603), and the Strategic Priority Research Program of Chinese Academy of Sciences No. XDB28000000.


\begin{thebibliography}{99}
\bibitem{Nielsen} M. A. Nielsen and I. L. Chuang, Quantum Computation and Quantum Information, Canbrudge University Press, Cambridge, 2000.
\bibitem{Shor} P. W. Shor, \textit{SIAM J. Comput.} \textbf{26}, 1484 (1997).
\bibitem{Grover} L. K. Grover, \textit{Phys. Rev. Lett.} \textbf{79}, 325 (1997).
\bibitem{Bennett} C. H. Bennett and G. Brassard, \textit{Theor. Comput. Sci.} \textbf{560}, 7 (2014).
\bibitem{Giovannetti} V. Giovannetti, S. Lloyd, and L. Maccone,  \textit{Science} \textbf{306}, 1330 (2004).
\bibitem{Giovannetti1} V. Giovannetti, S. Lloyd, and L. Maccone,  \textit{Nat. Photonics} \textbf{5}, 222 (2011).
\bibitem{Lambert} N. Lambert, Y.-N. Chen, Y.-C. Cheng, C.-M. Li,  G.-Y. Chen, and F. Nori, \textit{Nat. Phys.}  \textbf{9}, 10 (2013).
\bibitem{Aberg1} J. {\AA}berg, arXiv:quant-ph/0612146.
\bibitem{Levi} F. Levi and F. Mintert, \textit{New J. Phys.} \textbf{16}, 033007 (2014).
\bibitem{Baumgratz} T. Baumgratz, M. Cramer, and M. B. Plenio, \textit{Phys. Rev. Lett.} \textbf{113}, 140401 (2014).
\bibitem{Fan} M.-L. Hu, X. Hu, J.-C. Wang, Y. Peng, Y.-R. Zhang, and H. Fan, \textit{Phys. Rep.} \textbf{762-764}, 1 (2018).
\bibitem{Streltsov} A. Streltsov, G. Adesso, and M. B. Plenio, \textit{Rev. Mod. Phys.} \textbf{89}, 041003 (2017).
\bibitem{Horodecki} M. Horodecki and J. Oppenheim, \textit{Int. J. Mod. Phys. B} \textbf{27}, 1345019 (2013).
\bibitem{Chitambar} E. Chitambar and G. Gour, \textit{Rev. Mod. Phys.} \textbf{91}, 025001 (2019).
\bibitem{Brandao} F. G. S. L. Brand$\tilde{a}$o and G. Gour, \textit{Phys. Rev. Lett.} \textbf{115}, 070503 (2015).
\bibitem{Winter} A. Winter and D. Yang, \textit{Phys. Rev. Lett.} \textbf{116}, 120404 (2016).
\bibitem{Yadin} B. Yadin, J. Ma, D.  Girolami, M. Gu, and V. Vedral, \textit{Phys. Rev. X} \textbf{6}, 041028 (2016).
\bibitem{Lami1} L. Lami, B. Regula, and G. Adesso,\textit{Phys. Rev. Lett.} \textbf{122}, 150402 (2019).
\bibitem{Zhao} Q. Zhao, Y. Liu, X. Yuan, E. Chitambar, and A. Winter, \textit{IEEE Trans. Inf. Theory} \textbf{65}, 6441 (2019).
\bibitem{Lami} L. Lami, arXiv: 1902.02427.
\bibitem{Bu1} K. Bu, U. Singh, S.-M. Fei, A. K. Pati, and J. Wu, \textit{Phys. Rev. Lett.} \textbf{119}, 150405 (2017).
\bibitem{Regula} B. Regula, L. Lami, and A. Streltsov, \textit{Phys. Rev. A} \textbf{98}, 052329 (2018).
\bibitem{Regula1} B. Regula, K. Fang, X. Wang, and G. Adesso, \textit{Phys. Rev. Lett.} \textbf{121}, 010401 (2018).
\bibitem{Fang} K. Fang, X. Wang, L. Lami, B. Regula, and G. Adesso, \textit{Phys. Rev. Lett.} \textbf{121}, 070404 (2018).
\bibitem{Liu} C. L. Liu, Y. Q. Guo, and D. M. Tong, \textit{Phys. Rev. A} \textbf{96}, 062325 (2017).
\bibitem{Chitambar1} E. Chitambar, \textit{Phys. Rev. A} \textbf{97}, 050301(R) (2018).
\bibitem{Yuan} X. Yuan, H. Zhou, Z. Cao, and X. Ma, \textit{Phys. Rev.A} \textbf{92}, 022124 (2015).
\bibitem{Liu2} C. L. Liu and D. L. Zhou, \textit{Phys. Rev. Lett.} \textbf{123}, 070402 (2019).
\bibitem{Jonathan} D. Jonathan and M. B. Plenio,  \textit{Phys. Rev. Lett.} \textbf{83}, 3566 (1999).
\bibitem{Feng} Y. Feng, R. Duan, and M. Ying, \textit{IEEE Trans. Inf. Theory} \textbf{51}, 1090 (2005).
\bibitem{Turgut} S. Turgut, \textit{J. Phys. A: Math. Theor} \textbf{40}, 12185 (2007).
\bibitem{Klimesh} M. Klimesh, arXiv: 0709.3680.
\bibitem{Bu} K. Bu, U. Singh, and J. Wu, \textit{Phys. Rev. A} \textbf{93}, 042326 (2016).
\bibitem{Du1} S. Du, Z. Bai and Y. Guo, \textit{Phys. Rev. A} \textbf{91}, 052120 (2015).
\bibitem{Killoran} N. Killoran, F. E. S. Steinhoff, and M. B. Plenio,\textit{Phys. Rev. Lett.} \textbf{116}, 080402 (2016).
\bibitem{Du} S. Du, Z. Bai, and X. Qi, \textit{Quantum Inf. Comput.} \textbf{15}, 1307 (2015).
\bibitem{Vidal} G. Vidal,  \textit{Phys. Rev. Lett.} \textbf{83}, 1046 (1999).
\bibitem{Zhu} H. Zhu, Z. Ma, Z. Cao, S. M. Fei, and V. Vedral, \textit{Phys. Rev. A} \textbf{96} 032316 (2017).
\bibitem{Chitambar2} E. Chitambar and G. Gour, \textit{Phys. Rev. A} \textbf{94}, 052336 (2016).

\end{thebibliography}
\end{document}